\documentclass[11pt,a4paper]{article}
\pdfoutput=1
\usepackage{jcappub}

\usepackage{bm}
\usepackage{amsfonts}
\usepackage{subfigure}
\usepackage{placeins}
\usepackage{todonotes}

\newcommand{\be}{\begin{equation}}
\newcommand{\ee}{\end{equation}}
\newcommand{\bea}{\begin{eqnarray}}
\newcommand{\eea}{\end{eqnarray}}

\newcommand{\di}[1]{\text{#1}}

\usepackage{multirow}
\usepackage[normalem]{ulem}

\makeatletter
\newcommand{\semivast}{\bBigg@{2}}
\newcommand{\vast}{\bBigg@{3}}
\makeatother

\begin{document}



\title{The variance of the locally measured Hubble parameter explained with different estimators}

\author[a]{Io Odderskov,}
\author[a]{Steen Hannestad,}
\author[a,b]{Jacob Brandbyge}

\affiliation[a]{Department of Physics and Astronomy, Aarhus University, DK-8000 Aarhus C, Denmark}
\affiliation[b]{Centre for Star and Planet Formation, Niels Bohr Institute \& Natural History Museum of Denmark, University of Copenhagen, {\O}ster Voldgade 5-7, DK--1350 Copenhagen, Denmark}

\emailAdd{isho07@phys.au.dk, sth@phys.au.dk, jacobb@phys.au.dk}

\abstract{We study the expected variance of measurements of the Hubble constant, $H_0$, as calculated in either linear
perturbation theory or using non-linear velocity power spectra derived from $N$-body simulations. We compare the variance with that obtained by carrying out mock observations in the N-body simulations, and show that the estimator typically used for the local Hubble constant in studies based on perturbation theory is different from the one used in studies based on N-body simulations. The latter gives larger weight to distant sources, which explains why studies based on N-body simulations tend to obtain a smaller variance than that found from studies based on the power spectrum. Although both approaches result in a variance too small to explain the discrepancy between the value of $H_0$ from CMB measurements and the value measured in the local universe, these considerations are important in light of the percent determination of the Hubble constant in the local universe.}

\maketitle


\section{Introduction}

The Hubble parameter, $H$, measures the expansion rate of the Universe through the time derivative of the scale factor, $H \equiv \frac{\dot{a}}{a}$. Its current value, $H_0$, is one of the most fundamental cosmological observables. Historically the measurement of $H_0$ has been of great importance in e.g.\ establishing the age of the Universe. However, in recent years precision observations of the Cosmic Microwave Background (CMB) and large scale structure (LSS) have allowed for an indirect determination of $H_0$ with a precision exceeding that of current direct measurements. Intriguingly the value of $H_0$ inferred from CMB and LSS observations assuming a flat $\Lambda$CDM model \cite{Planck2015} 
\begin{equation}
H_0^{\rm Planck+BAO} =  67.6 \pm 0.6 \,\, {\rm km}\, {\rm s}^{-1}\, {\rm Mpc}^{-1}
\end{equation} 
is significantly lower than the value obtained from direct measurement in the local universe \cite{Riess2016}
\begin{equation}
H_0^{\rm local} = 73.24 \pm 1.74 \,\, {\rm km}\, {\rm s}^{-1}\, {\rm Mpc}^{-1}.
\label{eq:H0loc}
\end{equation}
There have been several suggestions for the origin of this discrepancy. It could be caused by systematic issues in either the value inferred from the CMB \cite{Clarkson2014,Planck2016intermediate} or in the local measurements \cite{Humphreys2013,Efstathiou2014} -- or both. However, as the considerations raised in these papers have been reexamined, and as more data is coming in, the discrepancy does not seem to be getting smaller \cite{Bonvin2015,Cardona2016,Riess2016}. Since the value of $H_0$ inferred from CMB and LSS observations is inferred using $\Lambda$CDM-like models, this could be an indication that the true cosmological model is different from $\Lambda$CDM.

An alternative possibility which has been discussed extensively is that the local universe is described by different parameters than the global ones. For example the presence of a large local underdensity -- a so-called Hubble bubble -- could lead to a significantly higher local value of $H_0$ relative to the global value \cite{Zehavi1998,Jha2007,Sinclair2010,Romano2016}.


A very interesting question is whether the variance predicted in local measurements of $H_0$ by $\Lambda$CDM models could actually account for the discrepancy. This has been studied in the literature several times using different approaches, resulting in somewhat differing conclusions. Calculations based on perturbation theory, such as 
\cite{Wang1998,Shi1998,Li2008,Marra2013,Ben-Dayan2014,Kaiser2015}, point to a variance in the local Hubble parameter of 2-3\%, for distributions of sources similar to the one used in the most recent measurements of the local Hubble parameter within $z = 0.1$. In contrast, studies based on N-body simulations result in a variance of less than 1\% in the same redshift range \cite{OdderskovHaugbolleHannestad,Wojtak2013,OdderskovKoksbangHannestad}. This discrepancy between the variance deduced from perturbation theory and from N-body simulations is the subject of this paper.

In this work, we estimate the variance in the local Hubble constant using both mock observations in a large N-body simulation and perturbation theory, and compare the results. In order to bridge the gap between the two methods, we also use a hybrid approach in which the variance is determined with perturbation theory but using the velocity power spectrum from the N-body simulation. In section \ref{sec:Mock_observations}, we describe the simulation used for the study, and explain how the mock observations are carried out. We also introduce two different estimators for $H_0$ -- as we will see, a large part of the difference between the results from studies based on respectively N-body simulations and perturbation theory stems from the fact that they use two different estimators. In section \ref{sec:H0_from_velPS}, we derive the equations which describe the variance in $H_0$ in linear perturbation theory for each of the two estimators. In section \ref{sec:velPS}, we present the velocity power spectrum from the N-body simulation. Our results are presented in section \ref{sec:Results}, and finally we conclude in section \ref{sec:Conclusion}.

\section{Mock observations in N-body simulations} 
\label{sec:Mock_observations}

\begin{figure}
	\centering
	\includegraphics{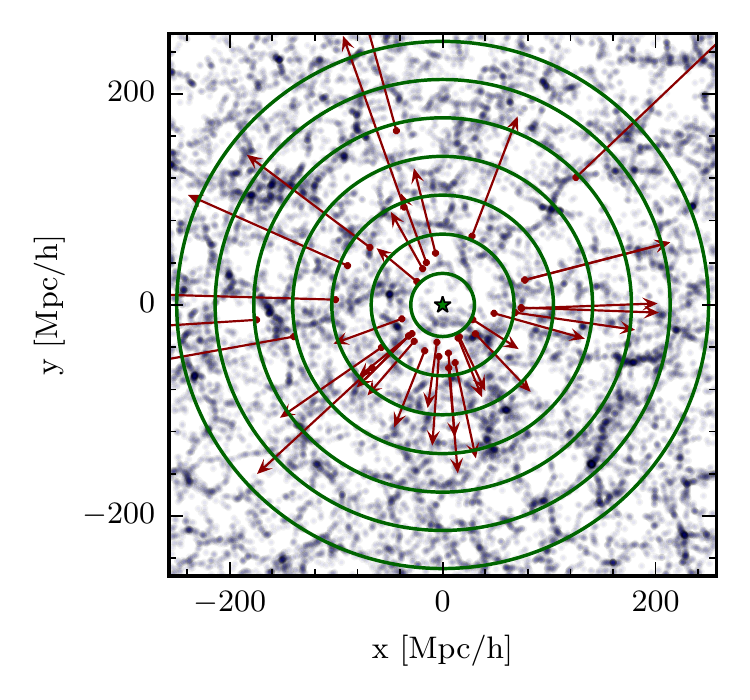}
	\includegraphics[width=0.47\textwidth]{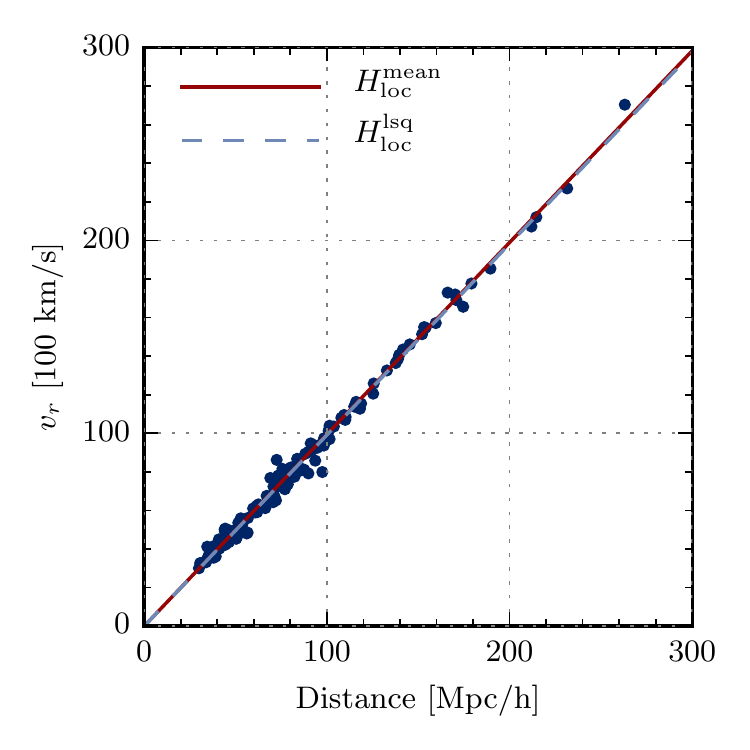}
	\hspace*{0.4cm}
	\includegraphics[width=0.47\textwidth]{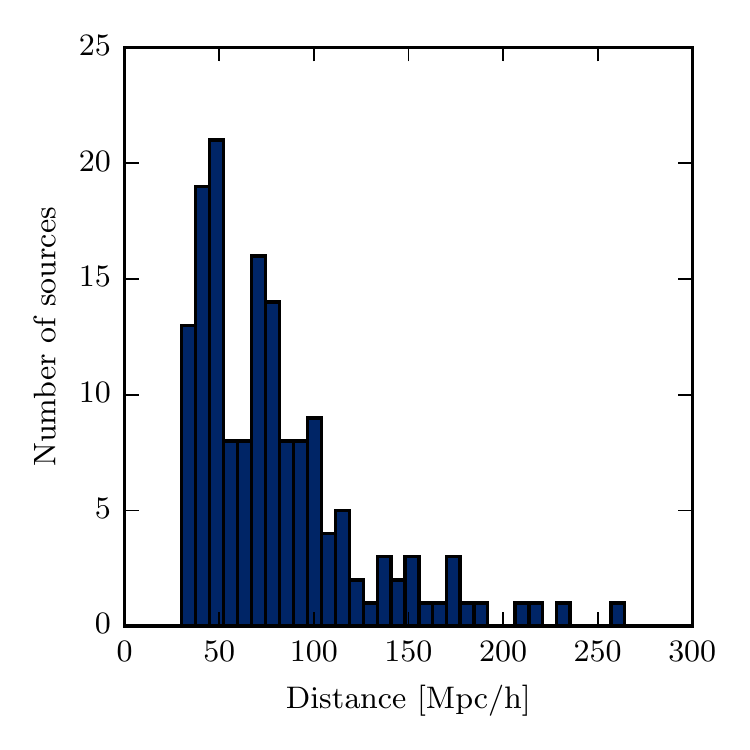}
	\includegraphics[width=0.47\textwidth]{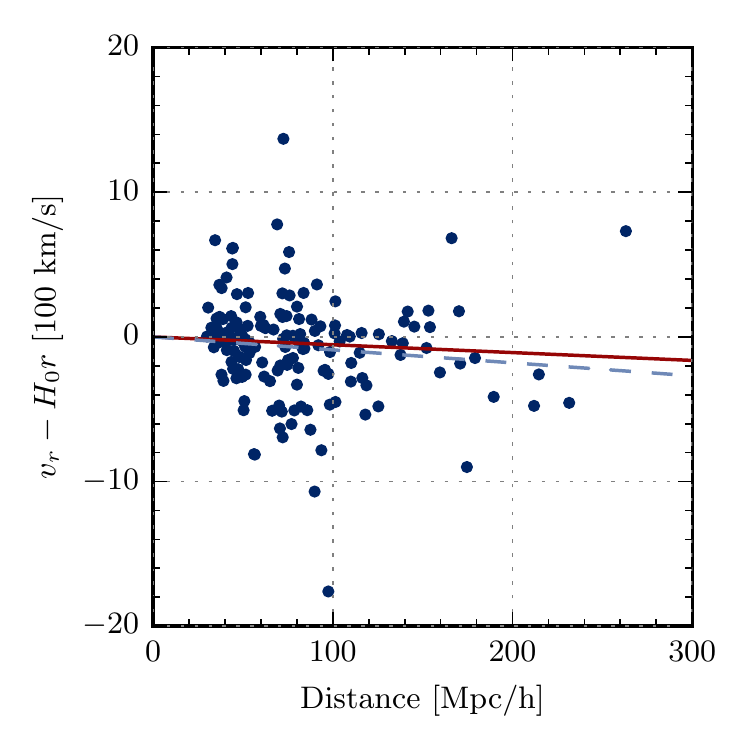}
	\caption{This figure illustrates the distribution of sources into distance bins, and shows the Hubble diagram along with two different estimates for $H_0$ for an example observer. In the \textbf{top left} figure, the example observer is marked with a green star, and the figure shows all the halos from a $10 \di{Mpc/h}$ thick slice centered on the observer in grey, and the observed sources in red along with their total radial velocities. The green circles show how these are distributed into distance bins -- only every 5th division is shown. In the \textbf{bottom left}, the distance distribution of sources is shown, corresponding to that of the CfA3+OLD collection of type Ia supernovae. In the \textbf{top right}, the resulting Hubble diagram is shown, illustrating two different estimators which differ in the weights assigned to sources at different distances -- these are denoted by $H_\textrm{loc}^\textrm{mean}$ and $H_\textrm{loc}^\textrm{lsq}$, and will be defined in section \ref{subsuc:estimators}. In the \textbf{bottom right}, the deviation from the background expansion of the simulation, given by $v_r = H_0 r$, is shown, to clarify how the two different estimates of $H_0$ deviate from each other.}
	\label{fig:example_observation}
\end{figure}

The most precise measurements of the local value of $H_0$  are based on measurements of the distances to galaxies, inferred from the observed luminosities of type Ia supernovae, together with the spectroscopic redshifts of their host galaxies which provide the radial velocities \cite{Riess2011,Riess2016}. An intuitive way to mimic such measurements is by carrying out mock observations in synthetic data sets, which can for example be generated from the halo catalogs from N-body simulations. In this section, we will describe such an analysis.

\subsection{Simulation}

The mock observations are based on an N-body simulation with cosmological parameters given by $(\Omega_b,\Omega_m) = (0.048, 0.26)$,  $(h,\sigma_8) = (0.68,0.85)$ and a spectral index of $n_s = 1$. The simulation is performed using a modified version of the GADGET-2 code \cite{Springel2005}, with initial conditions generated using a code described in \cite{Brandbyge2010} based on transfer functions computed using CAMB\footnote{\url{http://camb.info/}} \cite{Lewis2002}. The simulation is run in a box of side length $2048\,\, \di{Mpc/h}$, containing $1024^3$ dark matter particles. It is initiated at a redshift of $z=49$ and run until the present time. In this paper, we only consider the snapshot at $z=0$. We use the halo finder ROCKSTAR \cite{Behroozi2011} to generate the halo catalog from which sources for the local determination of $H_0$ will be selected.


\subsection{Observers and observations}

We carry out mock observations for 1,000 observers distributed throughout the simulation volume. Three different distance-distributions of sources are used for estimating the Hubble constant. The first two are chosen to be similar to the distributions used in actual observations; these are the CfA3+OLD collection of supernovae, which is close to the one used in \cite{Riess2011}, and the Union 2.1 collection, which covers the same redshift range as the one used in \cite{Riess2016}. In order to obtain source distributions correponding to each of these for the observers in the simulation, the volume around each observer is divided into bins of equal width -- as illustrated in figure \ref{fig:example_observation} -- and within each bin the number of sources dictated by the distance distribution is chosen among the halos in the bin. We also consider a third distribution, in which the sources are chosen randomly among all the halos in the simulation within a radius of $256\,\,\di{Mpc/h}$ of the individual observers; this corresponds to a top-hat distribution, and it peaks at high redshift, since the volumes of concentric shells grow with distance. Note that no measurements of the local value of the Hubble constant exists for such a distribution.  

Observers are placed at random positions throughout the simulation volume. In e.g. \cite{OdderskovHaugbolleHannestad,OdderskovKoksbangHannestad}, it was shown that the observer position causes a bias in the measured Hubble constant, since observers placed in massive halos will tend to measure a low value of $H_0$ due to local inflow, whereas observers placed at random positions -- which will tend to be located in voids as these take up a larger part of the simulation volume -- tend to measure a value of $H_0$ slightly higher than the true value. However, even though the choice of the observer positions causes a bias in the measured value of the Hubble constant, we find that it does not have a significant effect on its \textit{spread} for any of the redshift distributions of sources considered in this study. That is, placing the observers in massive halos or at random positions results in approximately the same variance in the locally measured Hubble constant. 

When the sources for the mock observations are chosen among the halos in the simulation, overdense regions are sampled more than voids. 
In order to determine how this affects the measured variance, we also carry out a set of mock observations in which the sources are chosen randomly from a regular grid onto which the peculiar velocities in the simulation has been interpolated. 

\subsection{Estimating the Hubble constant}
\label{subsuc:estimators}

For each observer, we determine the Hubble constant using two different approaches which result in slightly different values for $H_0$. In one, the local Hubble constant is obtained by calculating $v_r/r$ for each of the observed sources, and then taking the mean of all these values, i.e.
\begin{align}
H_\textrm{loc}^\textrm{mean} \equiv \langle v_r / r \rangle.
\end{align}
A different calculation of the local Hubble constant is obtained by using the least-squares estimator for the slope of the relationship $v_r = H_0 r$. This is given by 
\begin{align}
H_\textrm{loc}^\textrm{lsq} \equiv \langle v_r r \rangle / \langle r^2 \rangle.
\end{align} 
We can compare this to $H_\textrm{loc}^\textrm{mean}$ by rewriting $H_\textrm{loc}^\textrm{lsq}$ as
\begin{align}
H_\textrm{loc}^\textrm{lsq} \equiv \frac{\langle \frac{v_r}{r}r^2\rangle}{\langle r^2\rangle},
\label{eq:Hlsq} 
\end{align}  
which shows that the least squares estimate is equivalent to the mean of the individual slopes with each of these weighted by the square of the distance to the source.

The expectation values for $H_\textrm{loc}^\textrm{mean}$ and $H_\textrm{loc}^\textrm{lsq}$, i.e. the values obtained by taking the mean for a large number of observers in the simulation, are not exactly the same. Since $H_\textrm{loc}^\textrm{mean}$ gives relatively more weight to nearby sources, local flows affect this estimator more than $H_\textrm{loc}^\textrm{lsq}$. The previously mentioned tendency for the observers to measure a value of the local Hubble constant which is higher than the true value, due to the fact that they tend to be located in underdense regions, is therefore somewhat more pronounced for $H_\textrm{loc}^\textrm{mean}$ than for $H_\textrm{loc}^\textrm{lsq}$. The effect is largest for the CfA3+OLD distribution of sources, in the case where the sources are chosen among the halos in the simulation; in this case, the average values of the Hubble constant among the observers in the simulation for each of the two estimators are $\left\langle H_\textrm{loc}^\textrm{mean} \right\rangle = 100.41\,\, {\rm km}\, {\rm s}^{-1}\, {\rm Mpc}^{-1}\, {\rm h}$ and $\left\langle H_\textrm{loc}^\textrm{lsq} \right\rangle = 100.16\,\, {\rm km}\, {\rm s}^{-1}\, {\rm Mpc}^{-1}\, {\rm h}$.

In figure \ref{fig:example_observation}, the mock observation and subsequent determination of the Hubble constant is illustrated for an example observer. From the Hubble diagram for the specific observer it is difficult to make out the difference between the two estimates. To amplify the difference, we therefore also plot $v_r-H_0r$ against $r$. 


\section{Estimating the variance in $H_0$ from the velocity power spectrum}
\label{sec:H0_from_velPS}

The expected variance in the Hubble constant caused by over- and underdensitites in the Universe, and the resulting peculiar velocities, can be estimated from the power spectrum of density fluctuations. Below, we derive the expressions for the local variance of both $H_\textrm{loc}^\text{mean}$ and $H_\textrm{loc}^\textrm{lsq}$. We first derive the relationship between the velocity power spectrum and $H_\textrm{loc}^\mathrm{mean}$, as this is the simpler of the two cases.

\subsection{The local variation in $H_\textrm{loc}^\textrm{mean}$}

\begin{figure}
	\centering
	\includegraphics{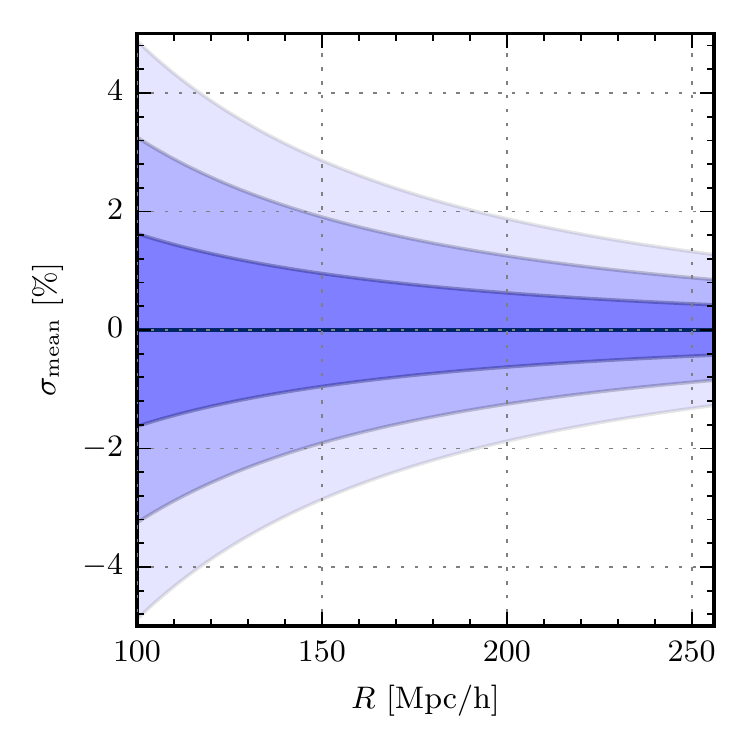}
	\includegraphics{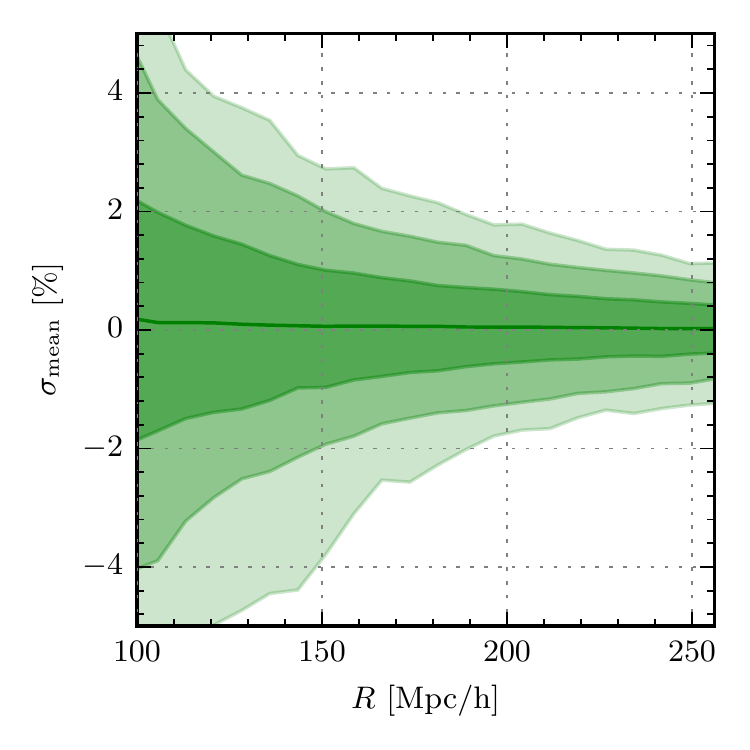}
	\caption{Deviations between $H_0$ and the local Hubble constant, calculated from respectively the linear velocity power spectrum from CAMB (\textbf{left}) and from mock observations in the N-body simulation (\textbf{right}). The full line shows the mean deviation of $H_\textrm{loc}^\textrm{mean}$ among observers in the box, and shadings indicate the 68\%, 95\% and 99\% confidence regions. For these plots, we have assumed a uniform -- i.e. a top-hat -- distribution of sources, with $R$ denoting the radius of the survey.}
	\label{fig:HubbleVariance}
\end{figure}	
	
Consider an observer at a comoving position $\textbf{r}_0$, making a local estimate of the Hubble constant based on observations of $N$ galaxies at positions $\textbf{r}_i$ and with peculiar velocities $\textbf{v}_i$. In this section it is assumed that the observer estimates the Hubble constant from the observed sources by taking the average of $v_r^i/r^i$, where $i$ indicates the individual sources - this is the approach used in  \cite{Shi1996,Wang1998,Shi1998}, and it is similar to that used in \cite{Marra2013,Ben-Dayan2014} (which, however, also incorporates non-linear and relativistic effects to a certain degree). This is equivalent to

\begin{align}
	H^\textrm{mean}_\textrm{loc} = H_0+\frac{1}{N}\sum_i \frac{\textbf{v}_i\cdot (\textbf{r}_i-\textbf{r}_0)}{|\textbf{r}_i-\textbf{r}_0|^2}.
\end{align}
We can replace the individual velocities $\textbf{v}_i$ by the velocity field, $\textbf{v}(\textbf{r})$, and the sum by an integral over a window function $W(\textbf{r}-\textbf{r}_0)$, which describes the distribution of sources:

\begin{align}
	H^\textrm{mean}_\textrm{loc} = H_0+ \int d^3\textbf{r} \frac{\textbf{v}(\textbf{r})\cdot (\textbf{r}-\textbf{r}_0)}{|\textbf{r}-\textbf{r}_0|^2} W(\textbf{r}-\textbf{r}_0).
		\label{eq:Hloc}
\end{align}	
The velocity field is generated by inhomogeneities in the density field, $\rho(\textbf{r})$, typically expressed through the over-density parameter $\delta(\textbf{r})=(\rho(\textbf{r})-\rho_0)/\rho_0$, where $\rho_0$ is the mean density. The velocity and the over-density fields can be expressed in terms of their Fourier transforms, i.e.

\begin{align}
	\textbf{v}(\textbf{r}) = \frac{1}{(2\pi)^3}\int d^3 \textbf{k}\, \textbf{v}_\textbf{k} e^{-i\textbf{k}\cdot \textbf{r}},
\end{align}
and likewise for $\delta(\textbf{r})$. In linear perturbation theory, the relationship between these fields is \cite{Peebles1980,Peebles1993} 
\begin{align}
	\textbf{v}_\textbf{k} = \frac{f H_0}{ik}\delta_\textbf{k} \hat{\textbf{k}},
	\label{eq:vk_from_dk}
\end{align}	
where $f$ is the growth function, which we approximate as $f = \Omega_m^{0.6}$, and $\hat{\textbf{k}}$ is a unit vector pointing in the direction of $\textbf{k}$. Inserting this in the integral of equation \ref{eq:Hloc}, it is found that the fractional deviation of the local Hubble constant from $H_0$ is given by
\begin{align}
	\frac{H^\textrm{mean}_\textrm{loc}-H_0}{H_0}=\frac{f}{(2\pi)^3}\int d^3 \textbf{k}\,\delta_\textbf{k} 
	e^{-i \textbf{k}\cdot \textbf{r}_0} \int d^3\, \textbf{r} \frac{\textbf{k}}{ik^2}\cdot \frac{\textbf{r}-\textbf{r}_0}{|\textbf{r}-\textbf{r}_0|^2}W(\textbf{r}-\textbf{r}_0)e^{-i\textbf{k}\cdot(\textbf{r}-\textbf{r}_0)}. 
	\label{eq:Hloc-H0}
\end{align}
The last integral represents the window function determining how density perturbations of different sizes affect the local Hubble constant. As a simple example, consider the case where $W$ is a top-hat window with radius $R$, i.e.

\begin{align}
	W_\textrm{top-hat}(\textbf{x})= \begin{cases}
	\frac{1}{4/3\pi R^3},\quad\quad&|\textbf{x}| \leq R,\\
	0,\quad\quad&|\textbf{x}| > R.
	\end{cases}
\end{align} 
In this case, the window function reduces to a relatively simple expression in Fourier-space \cite{Shi1996,Wang1998}. By making the substitution $\textbf{x}=\textbf{r}-\textbf{r}_0$, and choosing the coordinate system with the $z$-axis parallel to $\textbf{k}$, such that $\textbf{k}\cdot \textbf{x} = kx\cos(\theta)$, the integral can be evaluated as:
\begin{align}
	\mathcal{L}^\textrm{mean}_\textrm{top-hat}(kR)&\equiv \int d^3\, \textbf{x} \frac{\textbf{k}}{ik^2}\cdot\frac{\textbf{x}}{x^2}W_\textrm{top-hat}(x)e^{-i\textbf{k}\cdot \textbf{x}}\\
	&=\frac{1}{ik}\int_0^{R}dx \frac{3x}{4\pi R^3}\int_0^{2\pi}d\phi \int_0^{\pi}d\theta \sin(\theta)\cos(\theta) e^{-ikx\cos(\theta)}  \\
	&= \frac{3}{(kR)^3}\left(\sin (kR)- \int_0^{kR} dx\, \frac{\sin(x)}{x}\right).
\end{align}
For more realistic redshift distributions, such as those describing the CfA3+OLD and Union 2.1 collections of type Ia supernovae, an analytic expression for the window function cannot be obtained, and the integral has to be evaluated numerically.

In linear perturbation theory, the density field $\delta_k$, and consequently the deviations in the Hubble constant it produces, is a Gaussian random variable with mean $0$ and a variance, $\sigma^2_\textrm{mean} \equiv \left\langle \left(\frac{H^\textrm{mean}_\textrm{loc}-H_0}{H_0}\right)^2 \right\rangle$, that can be determined from equation \ref{eq:Hloc-H0} as
\begin{align}
	\sigma_\textrm{mean}^2 
	&= \frac{f^2}{(2\pi)^6}\int\int d^3\textbf{k}\,d^3\textbf{k}' \langle \delta_\textbf{k}\delta^*_{\textbf{k}'}\rangle \mathcal{L}^\textrm{mean}(kR)\mathcal{L}^\textrm{mean}(k'R)e^{-i(\textbf{k}-\textbf{k}')\cdot \textbf{r}_0}\\
	&=\frac{f^2}{2\pi^2R^2}\int_0^{\infty} dk\, P(k) \left[kR\mathcal{L}^\textrm{mean}(kR)\right]^2
	\label{eq:sigmamean}
\end{align}
where we have exploited that $\int d^3\textbf{x}e^{-i\textbf{k}\cdot\textbf{x}}=(2\pi)^3\delta^3_\textrm{D}(\textbf{k})$, with $\delta^3_\textrm{D}$ being the 3-dimensional Dirac delta function, and $P(k)=\langle |\delta_\textbf{k}|^2\rangle$ the matter power spectrum. It follows from equation \ref{eq:vk_from_dk} that the velocity power spectrum is related to the matter power spectrum as $P_v(k) = f^2 H_0^2 P(k)/k^2$. Hence, equation \ref{eq:sigmamean} for the variations in the Hubble constant can also be written in terms of the velocity power spectrum:
\begin{align}
	\sigma_\textrm{mean}^2 
	&=\frac{1}{2\pi^2R^2H_0^2}\int_0^{\infty} dk\, P_v(k) \left[k^2R\mathcal{L}^\textrm{mean}(kR)\right]^2.
	\label{eq:HubbleVariance}
\end{align}	
In figure \ref{fig:windows}, we show the window function, $\left[k^2R\mathcal{L}^\textrm{mean}(kR)\right]^2$ corresponding to the top-hat, the CfA3+OLD and the Union 2.1 redshift distributions.

Based on the velocity power spectrum from CAMB, we calculate and plot the variance in the Hubble constant calculated as in equation \ref{eq:HubbleVariance}. The resulting 68\%, 95\% and 99\% confidence regions as a function of the radius, $R$, of the top-hat window function are shown in figure \ref{fig:HubbleVariance}. We compare it to the equivalent result from the mock observations in the N-body simulation.

\subsection{The local variation in $H_\textrm{loc}^\textrm{lsq}$}

\begin{figure}
	\centering
	\includegraphics{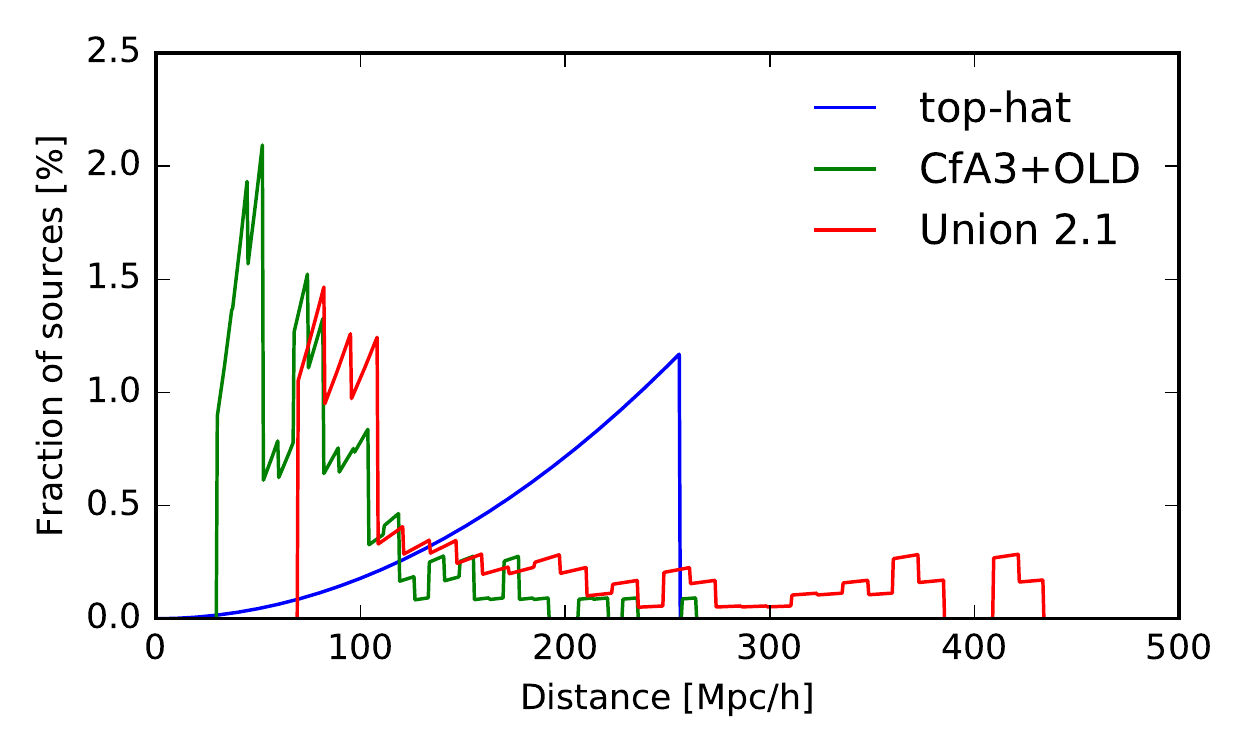}
	\includegraphics{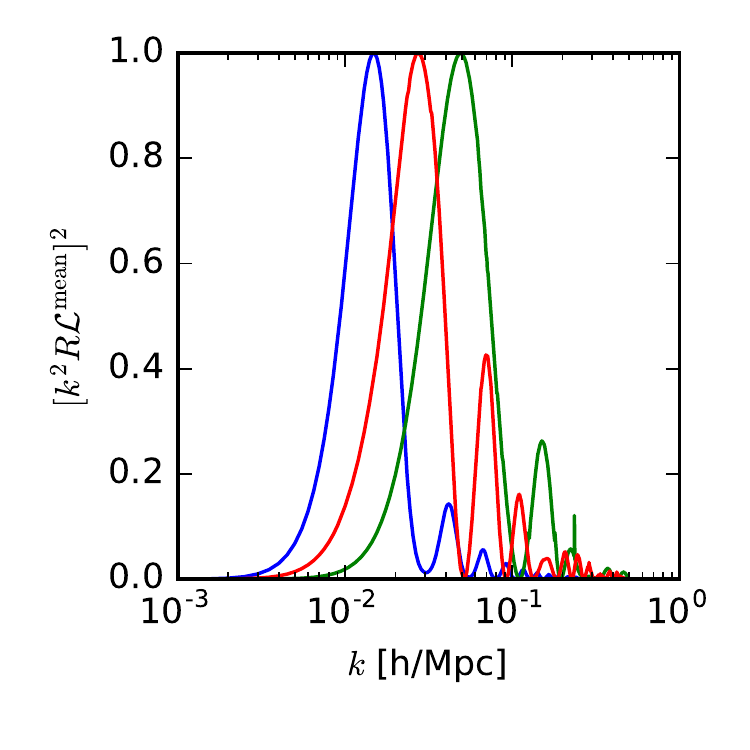}
	\includegraphics{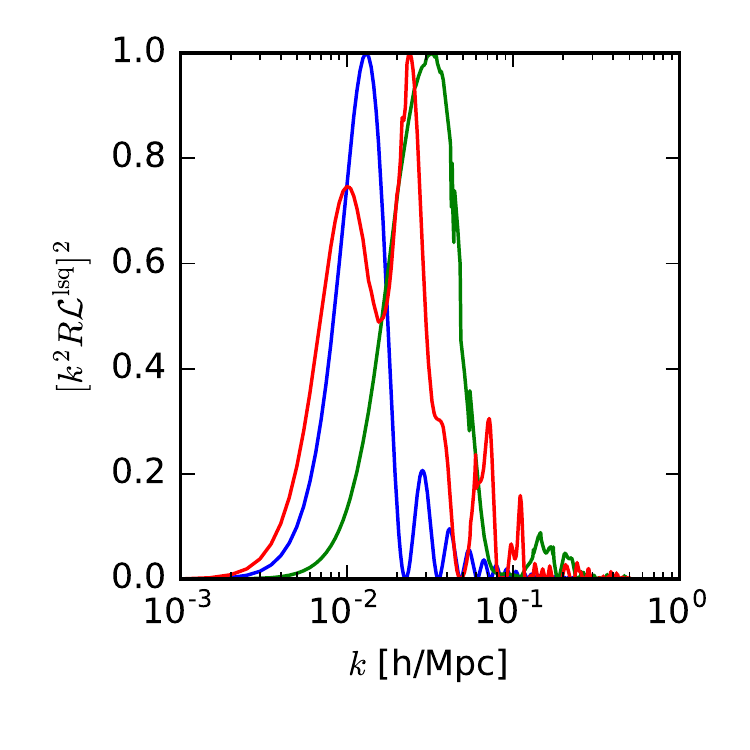}
	\caption{Distance distributions and window functions for the supernova samples considered in this study. Three different samples are used: a top-hat with radius $R=256\,\,\di{Mpc/h}$, corresponding to $z=0.087$, the CfA3+OLD type Ia supernova compilation, and Union 2.1 supernovae within $z=0.15$. The fractions of supernovae found at different redshifts for each of these distributions are shown in the \textbf{top} figure. On the \textbf{left}, we show the window functions for the variance in $H_\textrm{loc}^\textrm{mean}$, and on the \textbf{right} we show the window functions for the variance in $H_\textrm{loc}^\textrm{lsq}$. To facilitate the comparison, all the windows are normalized such that the maximal value is $1$.}
	\label{fig:windows}
\end{figure}

The calculation of the window function determining the local variations in the Hubble constant can be repeated for the least squares estimate of the Hubble constant, as given in equation \ref{eq:Hlsq}, by replacing equation \ref{eq:Hloc} by

\begin{align}
H^\textrm{lsq}_\textrm{loc} = H_0+ \frac{\int d^3\textbf{r}\, \textbf{v}(\textbf{r})\cdot (\textbf{r}-\textbf{r}_0)  W(\textbf{r}-\textbf{r}_0)}{\int\, d^3 \mathbf{r} |\mathbf{r}-\mathbf{r}_0|^2 W(\mathbf{r}-\mathbf{r}_0)  }.
\end{align}
This leads to a slightly different set of window functions, $\mathcal{L}^\textrm{lsq}$, given by
\begin{align}
\mathcal{L}^\textrm{lsq}(kR) \equiv 
\frac{\int d^3 \textbf{x}\, \textbf{k} \cdot \textbf{x} W(x)e^{-i\textbf{k}\cdot \textbf{x}}}
{i k^2 \int d^3 \mathbf{x}\, x^2 W(x)}.
\end{align}
The variance in the least squares estimates of $H_0$, $\sigma^2_\textrm{lsq} \equiv \left\langle \left(\frac{H^\textrm{lsq}_\textrm{loc}-H_0}{H_0}\right)^2 \right\rangle$, can then be found by simply replacing $\mathcal{L}^\textrm{mean}$ by $\mathcal{L}^\textrm{lsq}$ in equation \ref{eq:HubbleVariance}. We show the corresponding window functions, $\left[k^2R\mathcal{L}^\textrm{lsq}(kR)\right]^2$, in figure \ref{fig:windows}. Using the least squares estimates moves the window functions towards lower $k$-values, corresponding to larger scales, due to the larger weight given to distant sources. The top-hat window, in which the more distant sources already contribute much more than nearby sources, is almost unchanged.

We note that for each of the considered distributions, only modes with $0.005\,\, \di{h/Mpc}  \lesssim k \lesssim 0.1\,\, \di{h/Mpc}$ -- which are well resolved in the simulation used for this study -- contribute significantly to the variance in the Hubble constant.

\section{A hybrid approach: Velocity power spectrum from N-body simulations}
\label{sec:velPS}

We can replace the linear prediction for the velocity power spectrum by that measured in the N-body simulation in which the mock observations are carried out. By using the velocity power spectrum measured from the N-body simulation in the calculation of the variance in $H_\textrm{loc}^\textrm{mean}$ and $H_\textrm{loc}^\textrm{lsq}$, and comparing with the variance predicted by linear theory, we can determine how much the variance is affected by the non-linear evolution of the velocity field. In contrast to the matter power spectrum, the velocity power spectrum \textit{decreases} when non-linear effects of clustering are included. Therefore, the variance measured in N-body simulations can be expected to be lower than that determined in linear perturbation theory. However, this effect is only significant at relatively small scales ($k\gtrsim 0.1\,\, \di{h/Mpc}$), which only contribute very little to the variance in the measured Hubble constant for each of the distributions of sources studied here.

\subsection{The non-linear velocity power spectrum}

\begin{figure}
	\centering
	\hspace*{0.05cm}
	\includegraphics{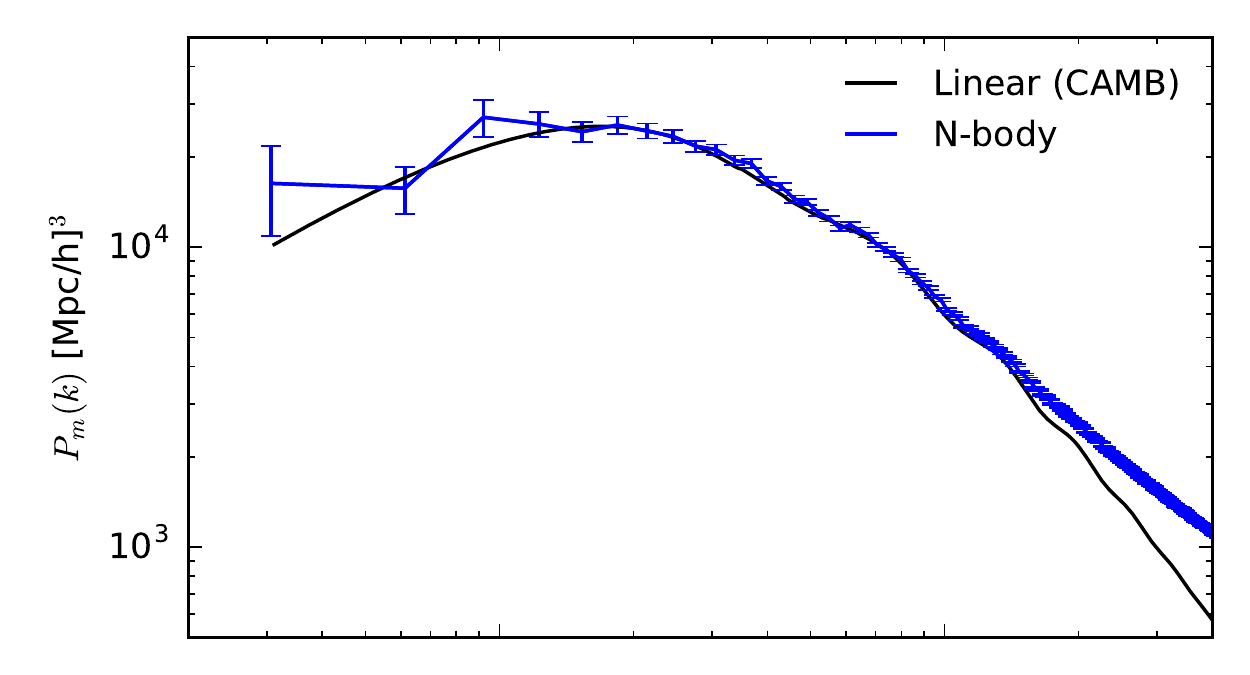}
	\includegraphics{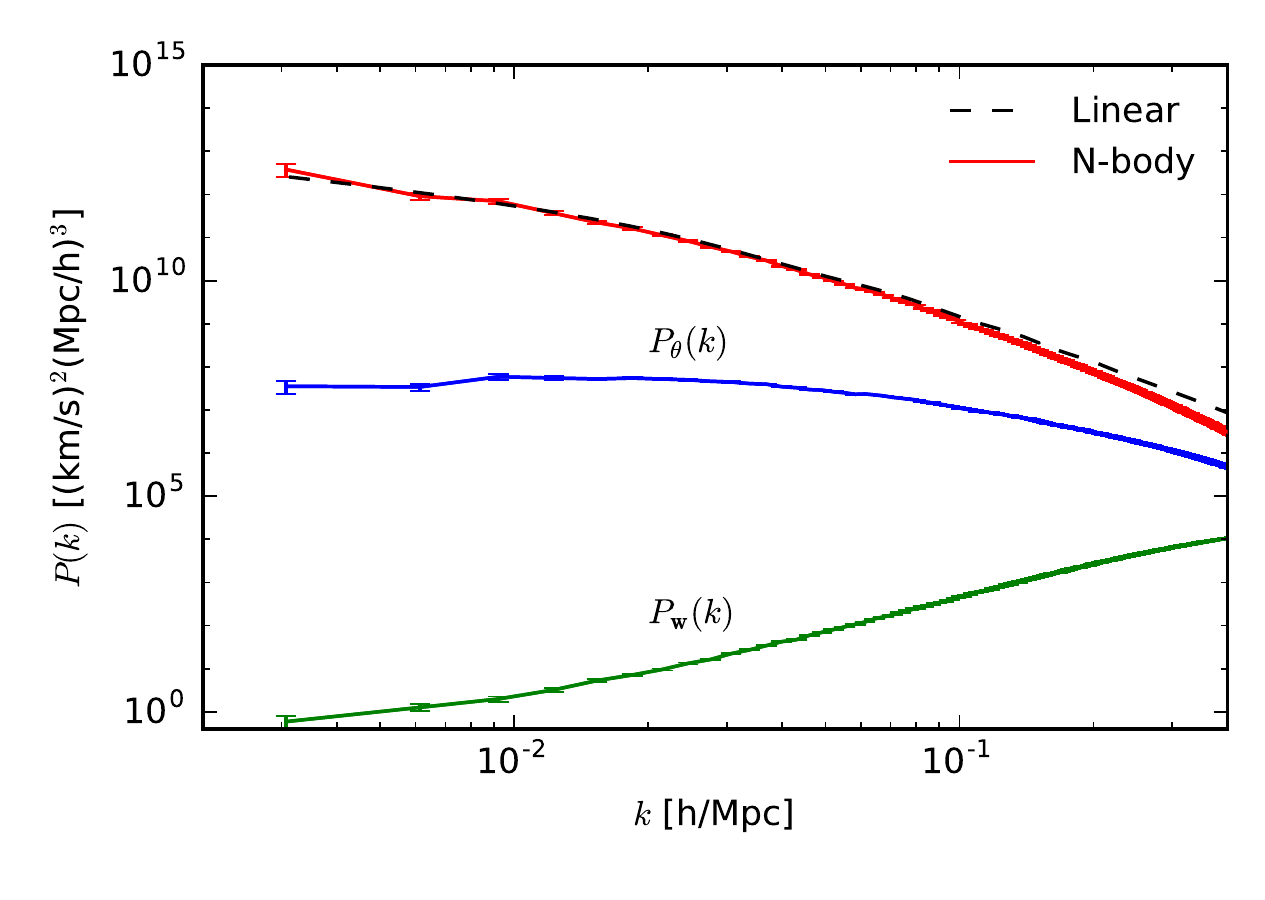}
	\caption{The matter power spectrum (\textbf{top}) and velocity power spectrum (\textbf{bottom}) from the N-body simulation, each compared to that from linear perturbation theory. The divergence and curl velocity power spectra are also shown. The curl power spectrum is not reliable when determined from a simulation with the resolution used here; however, this has no discernible effect on the results as the full velocity power spectrum is completely dominated by the divergence power spectrum on the relevant scales. All the errorbars are calculated as $P(k)/\sqrt{N}$, where $N$ is the number of modes in the given $k$-bin.}
	\label{fig:Nbody_powerspectra}
\end{figure}

From the velocity field we construct its divergence, $\theta = \nabla \cdot \mathbf{v}$, and curl, $\mathbf{w}=\nabla\times \mathbf{v}$. The velocity field can then be written as $\mathbf{v}=\nabla^{-2}\nabla\theta + (\nabla\times)^{-2}\nabla\times\mathbf{w}$. Fourier transforming we get $\mathbf{v}=\frac{i}{k^2}(-\mathbf{k} \theta  + \mathbf{k} \times\mathbf{w})$. The velocity power spectrum is then given by
\begin{align}
	P_v(k) = \langle \mathbf{v}(\mathbf{k}) \cdot \mathbf{v}^*(\mathbf{k})  \rangle,
\end{align}
with 
\begin{align}
	\mathbf{v}(\mathbf{k}) \cdot \mathbf{v}^*(\mathbf{k}) =& \frac{1}{k^4} \left[k^2 \theta\theta^* - \theta\mathbf{k}\cdot(\mathbf{k}\times\mathbf{w}^*)- \theta^*\mathbf{k}\cdot(\mathbf{k}\times\mathbf{w}) + (\mathbf{k}\times\mathbf{w})\cdot(\mathbf{k}\times\mathbf{w}^*)\right].
\end{align}
The second and third terms vanish through cyclic permutation, and the last term can be written as $k^2 \mathbf{w}\cdot \mathbf{w}^* -(\mathbf{k}\cdot\mathbf{w})(\mathbf{k}\cdot\mathbf{w}^*)$, where the latter term vanishes since the curl field is divergence free. Defining $P_\theta = \theta \theta^*$ and $P_w=\mathbf{w}\cdot\mathbf{w}^*$ we get
\begin{align}
	P_v = k^{-2} \left(P_\theta +P_w\right).
	\label{eq:Pv_sum}
\end{align}

The normalization of the divergence and velocity power spectra is given by the requirement that $\theta(\mathbf{k})=\delta(\mathbf{k}) H_0 f$ in the linear limit. This follows from the definition of the divergence power spectrum, since
\begin{align}
\theta(\mathbf{k})=\int d^3\textbf{r}\, \nabla \cdot \mathbf{v}(\mathbf{r}) e^{-i\mathbf{k}\cdot \mathbf{r}} = i\mathbf{k} \cdot \mathbf{v}(\mathbf{k}) = \delta(\mathbf{k}) H_0 f,
\end{align} 
where the last equality only holds in linear theory. 

All power spectra are normalized to the matter power spectrum from CAMB, which is shown in figure \ref{fig:Nbody_powerspectra} along with the matter power spectrum from the N-body simulation. Since for the velocity power spectrum, non-linear effects are non-negligible even at the largest scales in the N-body simulation at $z=0$, the normalization of $\theta$ is instead determined from a snapshot at $z=49$. In figure \ref{fig:Nbody_powerspectra}, the divergence, curl, and total velocity power spectra from the N-body simulation are shown, as well as the linear velocity power spectrum obtained from CAMB. The non-linear velocity power spectrum has been found by interpolating the N-body particle velocities to a $1024^3$ regular grid using the interpolation method of \cite{Monaghan1985}.

\section{Results}
\label{sec:Results}

\begin{center}
	\begin{table}
		\center 
		\begin{tabular}{lccc}
			\hline\hline
			\rule{0pt}{3ex}
			& Top-hat ($R=256\di{Mpc/h}$) 	& CfA3+OLD 	& Union 2.1 	\\
			\hline\\
			\multicolumn{4}{c}{------------------------ $\sigma_\mathrm{mean}$ ------------------------} \\ 
			\rule{0pt}{3ex} 
			Mock observations of halos & 0.49\% & 1.86\% & 0.81\% \\
			$\,\,$Mock observations of grid points & 0.47\% & 1.77\% & 0.77\% \\
			$\,\,$Linear velocity PS & 0.41\% & 1.72\% & 0.71\% 	 \\ 
			$\,\,$N-body velocity PS & 0.40\% & 1.62\% & 0.68\% 	\\
			\rule{0pt}{1ex}\\
			\multicolumn{4}{c}{------------------------ $\sigma_\mathrm{lsq}$ ------------------------} \\ 
			\rule{0pt}{3ex}
			Mock observations of halos & 0.39\% & 0.96\% & 0.30\% \\			
			$\,\,$Mock observations of grid points & 0.39\% & 0.93\% & 0.27\% \\						
			$\,\,$Linear velocity PS & 0.34\% & 0.86\% & 0.23\% 	\\
			$\,\,$N-body velocity PS & 0.33\% & 0.82\% & 0.22\% 	\\						
			\hline
		\end{tabular}
		\caption{This table gives the spread in each of the two estimates of the Hubble constant considered in this paper, $H_\textrm{loc}^\textrm{mean}$ (\textbf{top}) and $H_\textrm{loc}^\textrm{lsq}$ (\textbf{bottom}). In both cases, the spread is calculated based on three different distributions of sources, and using four different techniques: using mock observations with the observed sources chosen among the halos in the simulation; using mock observations with the sources chosen among the grid points onto which the velocity field has been interpolated; and using the velocity power spectrum (PS) obtained from respectively linear theory and from the N-body simulation.}
		\label{tab:results}
	\end{table}
\end{center}
\begin{figure}
	\centering
	\includegraphics{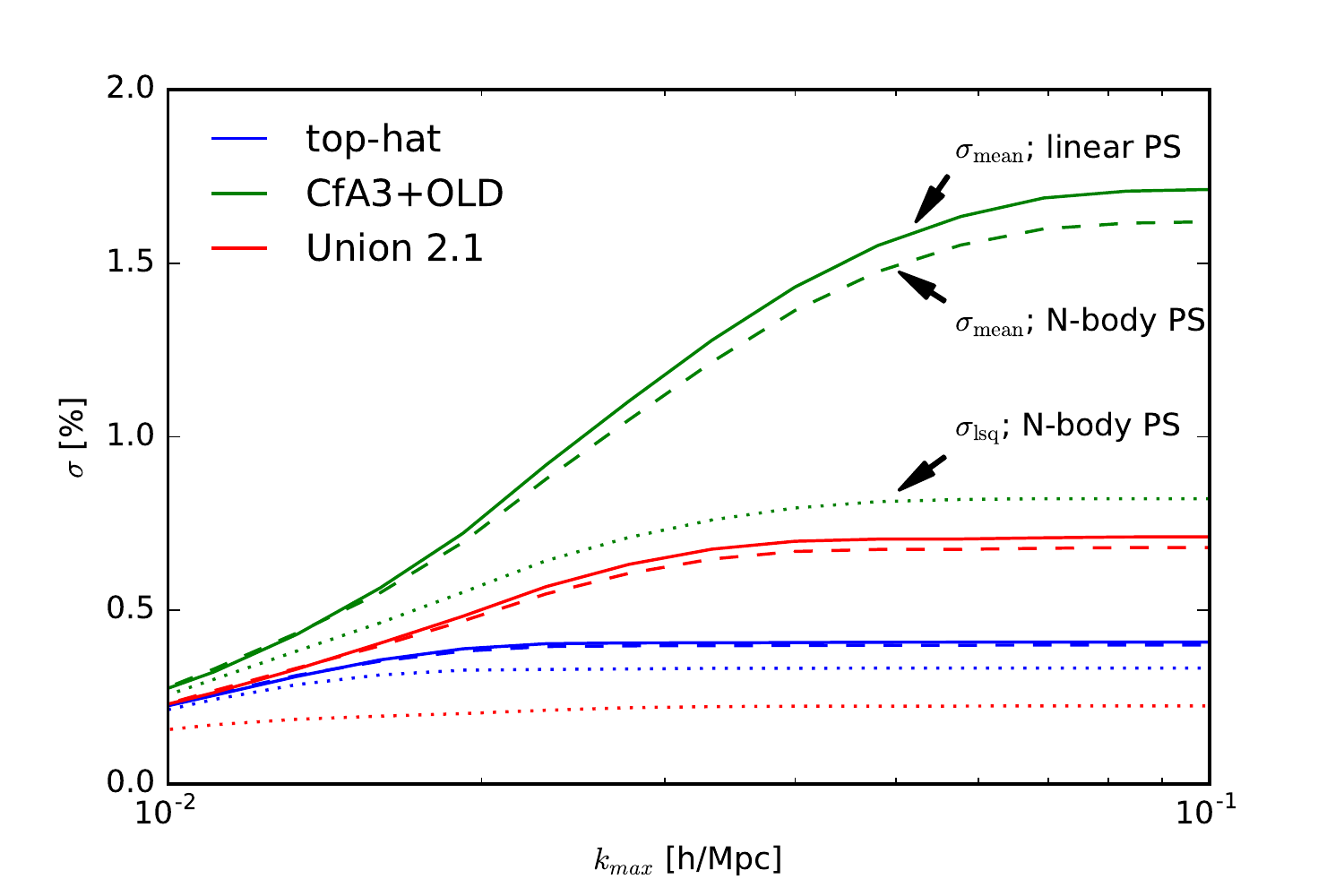}
	\caption{The variance -- calculated from the velocity power spectrum -- as a function of the highest $k$-values, i.e. smallest modes, included in the calculation. The different colors indicate different distributions of sources, as shown in the legend. Full lines show the results using the linear velocity power spectrum and the $H_\textrm{loc}^\textrm{mean}$ estimates for the Hubble constant; dashed lines show the result using the velocity power spectrum from the N-body simulation and $H_\textrm{loc}^\textrm{mean}$. We also show the result for the $H_\textrm{loc}^\textrm{lsq}$ estimator (with dotted lines), but only for the velocity power spectrum from the N-body simulation. The dotted lines are the ones that most closely correspond to what is measured when making mock observations in N-body simulations.}
	\label{fig:sigmaH0_against_kmax}
\end{figure}
In the previous sections, we presented two different estimators for $H_0$: $H_\textrm{loc}^\textrm{mean}$ and $H_\textrm{loc}^\textrm{lsq}$, and explained how their variances could be calculated from either mock observations or from the velocity power spectrum. In table \ref{tab:results}, we show the results corresponding to each of the window functions shown in figure \ref{fig:windows}. The calculation is both carried out for the velocity power spectrum predicted from linear perturbation theory, and for the velocity power spectrum measured in the N-body simulation. The mock observations are also carried out in two different ways, by selecting the observed sources among the halos in the simulation and among the grid points from which the velocity power spectrum is calculated. This results in a slightly smaller variance, as expected since peculiar velocities will tend to be smaller in voids than in overdense regions.

Even when the sources are chosen among the grid points, the spread in the Hubble constant found from the mock observations is 10-20\% higher than that deduced from the velocity power spectrum. 
The finite thickness of the bins and the randomness in the selection of sources described in section \ref{sec:Mock_observations} implies that there will be some variations between the distribution of sources used by the individual observers. Therefore, we expect the variance obtained from the mock observations to be slightly larger than that found from the calculations based on the power spectrum. By increasing the number of observed sources in the mock observations we have confirmed that the variance does indeed decrease by several percent when the velocity field is sampled more densely.     


In figure \ref{fig:sigmaH0_against_kmax} we show how the variance depends on the highest value of $k$, $k_\textrm{max}$, included in the analysis. As expected from the window functions, the value of $k_\textrm{max}$ has the largest impact on the variance associated with the CfA3+OLD supernova distribution, which peaks at the largest $k$-value, i.e. the smallest scales. It has nearly no effect on the top-hat window, for which only the largest scales in the simulation contribute significantly to the variance. \\

\section{Discussion and conclusions}
\label{sec:Conclusion}

We have calculated the variance in the local value of $H_0$ for two different estimators and three different distributions of sources. We have considered the differences caused by the non-linear evolution of the velocity power spectrum, as well as the significance of how the velocity field is sampled in the N-body simulation.  In the calculations based on perturbation theory, it was found that replacing the linear prediction for the velocity power spectrum by the fully non-linear velocity power spectrum measured in the N-body simulation only decreases the variance by 0.1\% or less. Similarly, changing the way the velocity field is sampled in the mock observations changes the variance by less than 0.1\%. 

In contrast, the estimator one uses to obtain the Hubble constant has a strong influence on the variance obtained, with the estimator typically used in studies based on the power spectrum, $H_\textrm{loc}^\textrm{mean}$, resulting in a variance about twice as large as the variance obtained using the estimator typically used in studies based on N-body simulations, $H_\textrm{loc}^\textrm{lsq}$. We have demonstrated that there is reasonable agreement between the variances obtained from perturbation theory and mock observations in N-body simulations if the same estimator is used in both cases.  

Neither of the two estimators considered in this study exactly correspond to how the Hubble constant is determined in actual observations. In real observations, greater weights are given to distant sources due to the fitting procedure, but this is most likely counteracted by the larger uncertainties in the redshifts and velocities of these sources compared to nearby ones. Moreover, actual surveys do not sample the whole sky, and hence have anisotropic window functions, which causes an additional variance, especially at large redshifts \cite{OdderskovHaugbolleHannestad,Wiegand2012}. An interesting and relevant extension of this study could take this approach even further, and calculate the exact window function corresponding to the estimator of $H_0$ used in modern determinations of the local Hubble constant, such as the one used in \cite{Riess2016}. As we have shown, this could be based either on mock observations or on the velocity power spectrum. Further, by basing the measurement of the velocity power spectrum on the halos in the simulation, each weighted by their type Ia supernova rate (which could e.g. be calculated from the star formation histories as in \cite{OdderskovHannestad}), the velocity power spectrum could be probed in the same way as it is in supernova surveys. This would allow for a thorough understanding of what exactly is measured in surveys of the local value of the Hubble constant, and how much this quantity can be expected to vary within any given cosmological model.

\section*{Acknowledgments}
We thank Dominik Schwarz for discussions that inspired this work and for useful comments. \\
\newline\noindent
Parts of the numerical work for this project have been done using computer resources from the Center for Scientific Computing Aarhus.\\
\newline\noindent
This work was supported by the Villum Foundation.

\FloatBarrier


\bibliographystyle{utcaps}

\bibliography{VelocityPS_and_H0}

\end{document}